\documentclass[%
 reprint,
 amsmath,amssymb,
 aps,
]{revtex4-1}

\usepackage{graphicx}
\usepackage{dcolumn}
\usepackage{bm}
\usepackage{color}
\newcommand{\tens}[1]{\mbox{\textbf{\textit{\textsf{#1}}}}}

\begin{document}

\preprint{APS/123-QED}

\title{Lateral Casimir-Polder forces by breaking time-reversal symmetry}
\author{Ricardo Oude Weernink}
\affiliation{%
Theoretisch-Physikalisches Institut, Friedrich-Schiller-Universit{\"a}t Jena
\\
 Fr{\"o}belstieg 1, 07743 Jena, Germany \\
}%

\author{Pablo Barcellona}
\email{pablo.barcellona@physik.uni-freiburg.de}
\affiliation{Physikalisches Institut, Albert-Ludwigs-Universit\"at
Freiburg, Hermann-Herder-Str. 3, 79104 Freiburg, Germany}

\author{Stefan Yoshi Buhmann}
\email{stefan.buhmann@physik.uni-freiburg.de}
\affiliation{Physikalisches Institut, Albert-Ludwigs-Universit\"at
Freiburg, Hermann-Herder-Str. 3, 79104 Freiburg, Germany}
\affiliation{Freiburg Institute for Advanced Studies,
Albert-Ludwigs-Universit\"at Freiburg, Albertstr. 19, 79104 Freiburg,
Germany}
\date{\today}

\begin{abstract}
\noindent
We examine the lateral Casimir-Polder force acting on a circular rotating emitter
near a dielectric plane surface. As the circular motion breaks time-reversal symmetry, 
 the  spontaneous emission in a direction parallel to the surface is in general anisotropic.
We show that a lateral force arises which can be interpreted as a recoil force because of this asymmetric emission.
The force is an
oscillating function of the distance between the emitter and the surface, and the lossy character of the dielectric strongly influences the results in the near-field regime. 
The force exhibits also a population-induced dynamics, decaying exponentially with respect to time on
timescales of the inverse of the spontaneous decay rate.
We propose that this effect could be detected measuring the velocity acquired by the emitter, following different cycles of excitation and thermalisation.
 Our results are expressed in terms of the Green's tensor and can therefore easily be applied to more complex geometries.

\begin{description}
\item[PACS numbers]
[42.50.Nn, 42.40.Lc, 12.20.Ds]
\end{description}
\end{abstract}

\maketitle

\section{\label{sec:level1}Introduction}
\noindent
Casimir-Polder (CP) forces are forces between atoms and magneto-dielectric bodies originating from the
quantum fluctuations of the electromagnetic field and the atomic charges \cite{CasPold,Cas}.
 Despite their generally low magnitude they
have to be taken into account when working on nanoscales which today is 
common in experiments as well as applications \cite{NanoTech}.

 Lateral Casimir forces are a relatively young topic, having first been
measured in 2002 \cite{chen}. They are characterised by
their direction which is parallel to the surface instead of the
usual normal direction, and have been suggested to facilitate contactless force transmissions \cite{ashourvan}.
Lateral forces are typically achieved by breaking the translational
symmetry of the surface, using for example periodically structured surfaces \cite{PerGrat,chen,chiu,rodrigues2,lambrecht} or corrugated surfaces \cite{CorrSur,dobrich,messina,contreras}.
A lateral force has also been realised by breaking the mirror symmetry, using chiral particles near a surface \cite{wang,hayat}. It is discriminatory since it pushes chiral particles with opposite handedness in opposite directions.

In this article we show that a lateral force can arise by breaking time-reversal symmetry via a rotating dipole which emits asymmetrically.
A dipole moment is created
with left-handed or right-handed circularly polarised light with spin parallel to the surface.
More specifically we excite a Cesium atom from the hyperfine ground-state $\left| {{6^2}{S_{1/2}},F = 4,{M_F} = 4} \right\rangle $ to the excited state $\left| 6^2 P_{3/2},F' = 5,M'_F = 5 \right\rangle $
using a resonant right-handed circularly polarised laser beam that propagates along the y-direction, creating a dipole moment rotating in the $x-z$ plane, where $\hat z$ is the direction normal to the surface.
The excitation of guided and radiation modes that propagate in the $+x$ and $-x$ directions may be expected to be asymmetric in this case \cite{LCPExp,xi}. The asymmetric emission of guided modes has been extensively  investigated in the  literature and relies on spin-orbit coupling of light mediated by a particle near a surface \cite{lin,rodriguez,luxmoore,neugebauer,mitsch,petersen,feber}.
The conservation of total momentum in the system in conjunction with the asymmetric emission suggests
the existence of a lateral force opposite to the direction of stronger emission.
This lateral force could be measured 
observing the asymmetric emission distribution.
A similar effect has been studied for the same atomic system close to an optical nanofiber \cite{LCPExp}.
It is the circular polarisation of the illuminating light that creates a rotating dipole-moment
which breaks time-reversal symmetry. Using linearly polarised light, the time-reversal symmetry is conserved and the force is purely normal to the surface.

Note that this system has been analysed previously by classical methods, studying the interaction of the rotating dipole with the reflected field \cite{rodriguez3}.
However the excited atom will  unavoidably decay to the ground state, for which the lateral force is forbidden by energy conservation. Hence the lateral force has a population-induced dynamics which can be captured only by quantum approaches to the atom-field coupling.
It is the aim of this article to investigate this effect.

The problem can be solved by expanding the electric field into guided and radiation modes. However here
we use a different approach which relies on the Green's tensor and permits to study 
geometries different from the planar configuration as well the impact of dissipation \cite{DF1,DF2}.

The article is organised as follows. In section II we study the asymmetric emission of a multilevel
circular emitter near a surface and investigate how the lateral force arises from this asymmetric emission.
We also develop a dynamical approach to the atom--field coupling which gives the same expression
for the lateral force.
In section III we consider a two-level Cesium atom near a surface to quantify the order of magnitude of the force and compare it with previous findings.

\section{\label{sec:level1}Lateral force from the asymmetric emission}
\begin{figure}[h!]
\centering
\includegraphics[width=7cm]{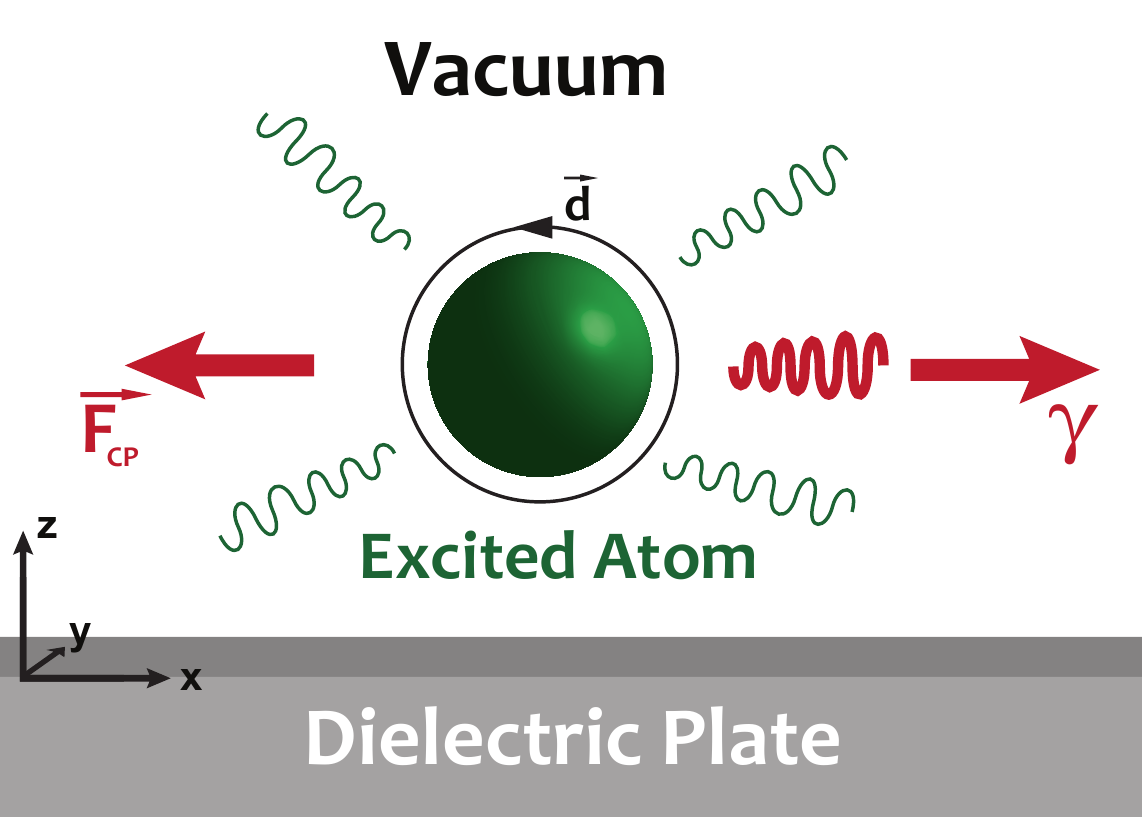}
\caption{Sketch of the system, expected lateral force and correlated asymmetric
decay due to rotational dielectric moment.}
\label{fig1}
\end{figure}

\noindent
We consider a multi-level atom, with atomic frequencies $\omega_{nk}=(E_n-E_k)/\hbar$, 
and dipole moments $\textbf{d}_{nk}$, near a homogeneous and isotropic dielectric plate (see Fig.~\ref{fig1}).
The atom is prepared in an incoherent superposition of energy eigenstates $\left| n \right\rangle $
with occupation $p_n(t)$ and is placed at
position $\textbf{r}_A =
z_{\operatorname{A}}\!\, \hat{\boldsymbol{z}}$ from the dielectric plate.
We assume the atomic eigenstate  $\left| n \right\rangle $ to exhibit a well-defined y-component of angular momentum. It is then a circular emitter
whose transitions matrix elements are not real, in general, but obey $\textbf{d}_{kn}=\textbf{d}_{nk}^*$ \cite{DF1}.
\\
We first derive the lateral force using a dynamical approach to the atom--field coupling.
According to the dynamical approach the Casimir-Polder force between
the atom and the surface reads \cite{buhmann,DF2}:
\begin{multline}
\textbf{F} (\textbf{r}_{\operatorname{A}},t)= \frac{\mu _0}{\pi }\mathop {\lim }\limits_{\varepsilon  \to 0^ + } \sum\limits_{n} p_n(t) \sum\limits_k \int\limits_0^\infty  \text{d}\omega \omega ^2  \frac{1}{\omega  - \omega _{nk} - \text{i}\varepsilon } \\
\times \nabla \textbf{d}_{nk} \cdot \text{Im} \tens{G}\left( \textbf{r},\textbf{r}_{\operatorname{A}},\omega  \right) \cdot \textbf{d}_{kn} \Big|_{\textbf{r} = \textbf{r}_{\operatorname{A}}+ \text{c.c.}}\,.
\end{multline}
This integral can be rotated to the imaginary axis leading to the following non-resonant and resonant contributions:
\begin{multline}
\textbf{F}^{\text{nr}} (\textbf{r}_{\operatorname{A}},t)=   \frac{\mu _0}{\pi} \sum\limits_{n} p_n(t) \sum\limits_{k }\int\limits_0^\infty  \text{d}\xi  \xi^2 \Bigg( \frac{1}{\omega _{nk} - \text{i}\xi } \\
+ \frac{1}{\omega _{nk} + \text{i}\xi } \Bigg)
 \text{Re} \Big\{ \nabla \textbf{d}_{nk} \cdot \tens{G}\left( \textbf{r},\textbf{r}_{\operatorname{A}},\text{i} \xi \right) \cdot \textbf{d}_{kn} \Big\} \Big|_{\textbf{r} = \textbf{r}_{\operatorname{A}}}\,,
 \end{multline}
 \begin{multline}
\textbf{F}^{\text{r}} (\textbf{r}_{\operatorname{A}},t)=   2\mu _0 \sum\limits_{n} p_n(t) \sum\limits_{k < n} \omega _{nk}^2\\
\times \text{Re} \Big\{ \nabla \textbf{d}_{nk} \cdot \tens{G}\left( \textbf{r},\textbf{r}_{\operatorname{A}},\omega _{nk} \right) \cdot \textbf{d}_{kn} \Big\} \Big|_{\textbf{r} = \textbf{r}_{\operatorname{A}}}\,.
\end{multline}
We first note that the lateral component of the non-resonant contribution vanishes, since $\partial _x \tens{G}^{(1)}\left( \textbf{r},\textbf{r}_{\operatorname{A}},\text{i} \xi \right)\big|_{\textbf{r} = \textbf{r}_{\operatorname{A}}}$ is an antisymmetric tensor and $\tens{G}^{(1)}\left( \textbf{r},\textbf{r}_{\operatorname{A}},\text{i} \xi \right)$ is real.
In fact the non-resonant term survives also for ground-state atoms, where the existence of a lateral
force is forbidden by energy conservation: if it existed, one could accelerate the atom along the surface, 
leaving the atom and the field in the stationary ground-states.
A lateral force can arise only from the resonant contribution, which describes the atom recoil because
of the asymmetric emission of the photon as we will show later. 

Most importantly the force shows a population induced dynamics. For example if the atom
is a two-level system initially excited in the state $\left| 1 \right\rangle $ the populations of the ground-state and excited-state satisfy the rate equations:
\begin{align}\nonumber
\dot p_0 &= \Gamma\, p_1\,,\\
\dot p_1 &=  - \Gamma\, p_1\,,
\end{align}
where $\Gamma$ is the spontaneous decay rate for the transition $1 \to 0$ (0 is the ground-state).
The solutions, subject to the initial conditions $p_0(0)=0$ and $p_1(0)=1$, are:
\begin{align}\nonumber
p_0\left( t \right) = 1 - \text{e}^{ - \Gamma t}\,,\\
p_1\left( t \right) =  \text{e}^{ - \Gamma t}\,.
\end{align}
The lateral force for this two-level system hence reads:
\begin{multline}
\operatorname{F}_x (\textbf{r}_{\operatorname{A}},t)
=2\mu _0 \text{e}^{ - \Gamma t} \omega _{10}^2 \\
\times \text{Re} \Big\{ \textbf{d}_{10} \cdot \partial _x \tens{G}^{(1)}\left( \textbf{r},\textbf{r}_{\operatorname{A}},\omega _{10} \right) \cdot \textbf{d}_{01} \Big\} \Big|_{\textbf{r} = \textbf{r}_{\operatorname{A}}} \,,
\label{eqn4}
\end{multline}
and decays exponentially with respect to time. Because of this force the particle will acquire some velocity in the lateral direction. 
Supposing that the initial velocity is zero, the mean-velocity acquired reads:
\begin{equation}
v = \int\limits_0^\infty  \text{d}t\frac{{\operatorname{F}}_x (\textbf{r}_{\operatorname{A}},t)}{m}  = \frac{{\operatorname{F}}_x (\textbf{r}_{\operatorname{A}},0)}{m \Gamma}\,.
\label{vel}
\end{equation}

We next  investigate the asymmetric emission of the circular emitter and how the lateral force arises from this asymmetry.
The total emission rate $\Gamma$ is the sum of the free-space emission rate $\Gamma^{(0)}$ and the surface-assisted emission rate $\Gamma^{(1)}$
where the photon is reflected by the dielectric surface. If the atom is prepared in the eigenstate $\left| n \right\rangle $ it will decay to lower lying energy levels, and the decay rate can be expressed in terms of the scattering Green's tensor $\tens{G}^{(1)}$ \cite{buhmann,DF2,sebastian10}:
\begin{multline}
\Gamma^{(1)}_n ({\operatorname{z}}_{\operatorname{A}})  = \frac{2\mu _0}{\hbar }\sum\limits_{k < n} \omega _{nk}^2 \\
\times  \text{Im} \Big\{ \textbf{d}_{nk} \cdot \tens{G}^{(1)}\left( \textbf{r}_{\operatorname{A}},\textbf{r}_{\operatorname{A}},\omega _{nk} \right) \cdot \textbf{d}_{kn} \Big\} \,,
\end{multline}
where we have used the property $\mathbf{d}_{nk} \cdot \text{Im} \tens{G}\left( \mathbf{r},\mathbf{r},\omega  \right) \cdot \mathbf{d}_{kn} = \text{Im} \left\{ \mathbf{d}_{nk} \cdot \tens{G}\left( \mathbf{r},\mathbf{r},\omega  \right) \cdot \mathbf{d}_{kn} \right\}$ .

The scattering Green's tensor for a dielectric reads ($z>0,z'>0$) \cite{DF1,DF2}:
\begin{align}\nonumber
\tens{G}^{\left( 1 \right)}\left( \textbf{r},\textbf{r}',\omega  \right) =&\int\limits_0^{2\pi } \text{d}\varphi  \int\limits_0^\infty  \text{d} k^\parallel k^\parallel   \tens{G}^{\left( 1 \right)}\left( \textbf{r},\textbf{r}',\omega ,\textbf{k}^\parallel  \right)\,,\\ \nonumber
\tens{G}^{\left( 1 \right)}\left( \textbf{r},\textbf{r}',\omega ,\textbf{k}^\parallel  \right)=&\frac{\text{i}}{8\pi ^2}\frac{1}{k^\bot } \text{e}^{\text{i} \textbf{k}^\parallel  \cdot \left( \textbf{r} - \textbf{r}' \right)} \text{e}^{\text{i} k^ \bot \left( z + z' \right)}\\
&\times \sum\limits_{\sigma  = s,p} r_\sigma  \textbf{e}_{\sigma _ + }\textbf{e}_{\sigma _ - } \,,
\label{eqn1}
\end{align}
where $\textbf{k}^\parallel = k^\parallel \left( \cos \phi ,\sin \phi ,0 \right)$ is wave vector parallel to the surface and $k^\bot  = \sqrt { \omega ^2/c^2-k^{\parallel 2}} $, $k_m^\bot  = \sqrt { \epsilon(\omega)\omega ^2/c^2-k^{\parallel 2}} $ are the perpendicular components of the wave-vector
in vacuum and in the dielectric plate. The polarisation vectors and Fresnel reflection coefficients for $s$-polarised and $p$-polarised waves read:
\begin{align}\nonumber
\mathbf{e}_{s \pm } =& \left( \sin \varphi , - \cos \varphi ,0 \right),\\ \nonumber
\mathbf{e}_{p \pm } =& \frac{c}{\omega }\left(  \mp k ^ \bot \cos \varphi , \mp k ^ \bot \sin \varphi , k^\parallel  \right), \\\nonumber
r_s =& \frac{k^ \bot  - k_m^ \bot }{k^ \bot  + k_m^ \bot }, \\ \nonumber
r_p =& \frac{\epsilon (\omega) k^ \bot  - k_m^ \bot }{\epsilon (\omega) k^ \bot  + k_m^ \bot }.
\end{align}
Using Eq. (\ref{eqn1}) the assisted rate reads:
\begin{equation}
\Gamma^{(1)}_n ({\operatorname{z}}_{\operatorname{A}}) =\int\limits_0^{2\pi } \text{d}\varphi  \int\limits_0^\infty  \text{d} k^\parallel k^\parallel   \gamma_n \left({\operatorname{z}}_{\operatorname{A}},  \textbf{k}^\parallel  \right)
\label{eqn50}\,,
\end{equation}
where $\gamma_n \left(z_A, \textbf{k}^\parallel  \right)$ is the emission rate density:
\begin{multline}
\gamma_n \left( {\operatorname{z}}_{\operatorname{A}}, \textbf{k}^\parallel  \right) = \frac{2\mu _0}{\hbar } \sum\limits_{k < n} \omega _{nk}^2 \\
\times \text{Im} \Big\{ \textbf{d}_{nk} \cdot \tens{G}^{(1)}\left( \textbf{r}_{\operatorname{A}},\textbf{r}_{\operatorname{A}},\omega _{nk},\textbf{k}^\parallel   \right) \cdot \textbf{d}_{kn} \Big\} \,.
\label{gamma4}
\end{multline}
A lateral force may result from an unbalanced spontaneous emission into the $+x$ and $-x$ directions.
According to the conservation of the total momentum the force is opposite to the momentum of the emitted
photon. If the atom is prepared in an incoherent superposition of energy eigenstates  $\left| n \right\rangle $ with population $p_n$ the force reads: 
\begin{multline}
{\operatorname{F}}_x (z_A,t)=  - \int\limits_0^{2\pi } \text{d}\varphi  \int\limits_0^\infty  \text{d} k^\parallel k^\parallel   \hbar k_x \sum\limits_{n} p_n \gamma_n \left({\operatorname{z}}_{\operatorname{A}}, \textbf{k}^\parallel  \right)\\
=2\mu _0\sum\limits_{n} p_n(t)  \sum\limits_{k < n} \omega _{nk}^2 \\
\times \text{Re} \Big\{ \textbf{d}_{nk} \cdot \partial _x \tens{G}^{(1)}\left( \textbf{r},\textbf{r}_{\operatorname{A}},\omega _{nk} \right) \cdot \textbf{d}_{kn} \Big\} \Big|_{\textbf{r} = \textbf{r}_{\operatorname{A}}} \,,
\label{eqn2}
\end{multline}
where we have used the relation $\text{Im} \left( \text{i}x \right) =  - \text{Re} x$.
 The lateral force is associated with the recoil of the atom because
of asymmetric emission. It vanishes if the atom is in the ground-state or
if the atomic dipole moments are real since $\partial _x \tens{G}^{(1)}\left( \textbf{r},\textbf{r}_{\operatorname{A}},\omega _{nk} \right)\big|_{\textbf{r} = \textbf{r}_{\operatorname{A}}}$ is an antisymmetric tensor.
Note that the bulk part of the Green's tensor gives no contribution to the lateral force since in the absence
of the dielectric plate the emission is obviously symmetric.

\section{\label{sec:level1}Application: Cesium atom}
The aim of this section is to investigate the lateral force for a Cesium atom near a dielectric plate
and compare the orders of magnitude of the force to previous works in literature 
using Cesium atoms near a optical nanofiber \cite{LCPExp}.

The Cesium atom in its ground-state can be excited to the 
excited state $\left| 6^2 P_{3/2},F' = 5,M'_F = 5 \right\rangle $ by using a right-handed circularly polarised laser beam. Since the beam is propagating along the $y$ direction the resulting electric dipole moment is rotating in
the $x-z$ plane:
\begin{align}
\boldsymbol{{\operatorname{d}}}_{10} = d \left(\text{i},\; 0, \; 
1 \right) \,,
\end{align}
where $1$ denotes the excited state and $0$ the ground-state.
The magnitude of the dipole moment is $d=1.9 \times 10^{-29} \mathrm{Cm}$ and the wavelength of the emitted photon is $\lambda_{10}= 852\, \mathrm{nm}$ \cite{LCPExp}.
Furthermore  $\boldsymbol{{\operatorname{d}}}_{01}=\boldsymbol{{\operatorname{d}}}_{10}^*$.
The emitted photon carries the momentum and is responsible for
the lateral force. 
There is only one decay channel to the ground state with the emission of a $\sigma^+$ photon, the atom hence can be treated as a effective two-level system.

Substituting the Green's tensor (\ref{eqn1}) into (\ref{eqn4}) and performing the trivial angular
integrals we find:
\begin{equation}
{\operatorname{F}}_x(\textbf{r}_{\operatorname{A}},t)
=  -\frac{\text{e}^{-\Gamma t}{\operatorname{d}}^2}{2\pi \varepsilon _0} \text{Im} \left\{ \int\limits_0^\infty  \text{d} k^\parallel k^{\parallel 3} \text{e}^{2\text{i} k^ \bot  {\operatorname{z}}_{\operatorname{A}}} r_p \right\}\,.
\label{eqn3}
\end{equation}

In the limit of short times the force does not vanish in the classical limit $\hbar \to 0$, showing that it could be described by classical
methods \cite{rodriguez3}. Only the $p$-polarised components of the field affect the force, since the $s$-polarised components lead to an emission equal in the $+x$ and $-x$ half spaces.

\begin{figure}[h!]
\centering
\includegraphics[width=7cm]{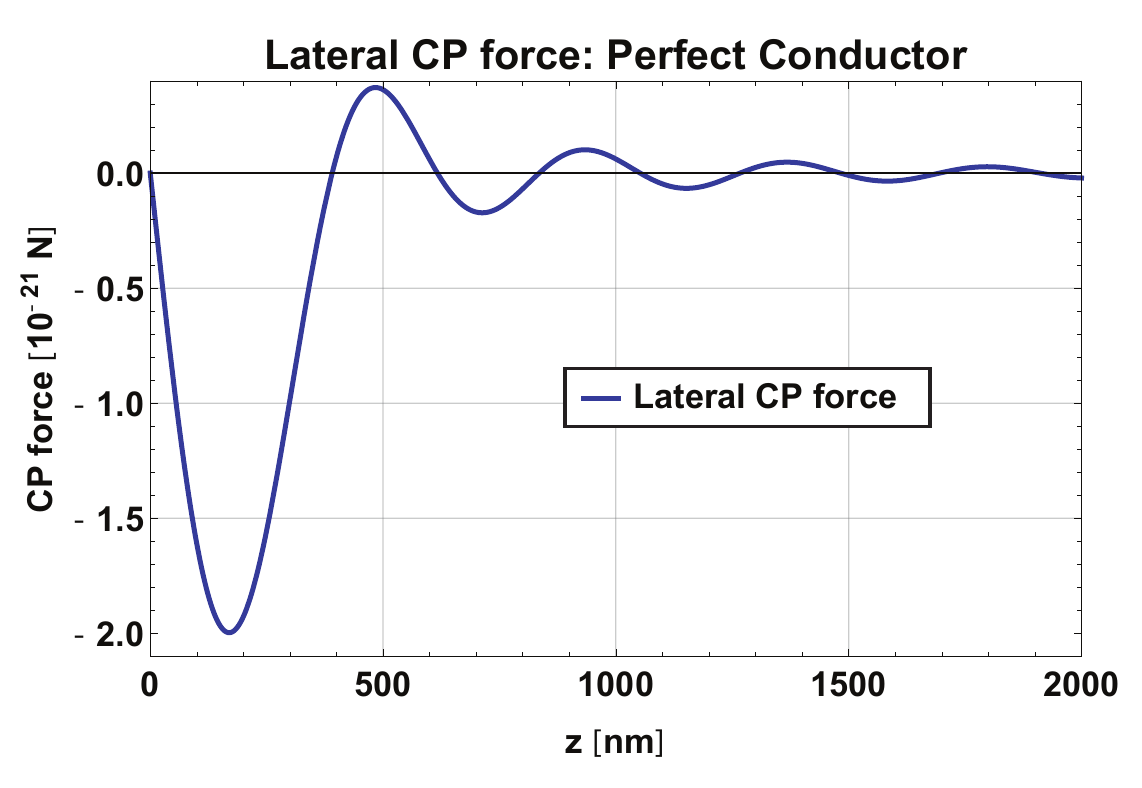}
\caption{LCP force for a perfectly conducting half space.}
\label{Fig2}
\end{figure}

The first interesting special case is a perfect conducting body, described by a unity reflection coefficient of p-polarised waves $r_p=1$. In this case the integration over the parallel component of the wave vector can be performed analytically for any given distance:
\begin{multline}
{\operatorname{F}}_x^{\text{PC}}\left(\textbf{r}_{\operatorname{A}},t \right) = \text{e}^{-\Gamma t} \Big\{ \frac{3{\operatorname{d}}^2}{4\varepsilon _0 \lambda _{10}}\frac{1}{{\operatorname{z}}_{\operatorname{A}}^3} \cos \left( 4\pi \frac{{\operatorname{z}}_{\operatorname{A}}}{\lambda _{10}} \right) \\
+ \frac {{\operatorname{d}}^2}{\varepsilon_0} \left( 
\frac {\pi}{\lambda_{10}^2} \frac {1}{{\operatorname{z}}_{\operatorname{A}}^2}
- \frac {3}{16\pi} \frac {1}{{\operatorname{z}}_{\operatorname{A}}^4}\right)
\sin\left(4\pi \frac {{\operatorname{z}}_{\operatorname{A}}}{\lambda_{10}} \right) \Big\} \;.
\end{multline}\\
where $k=\omega_{10}/c$ is the transition wave-vector.
This is depicted in Fig. (\ref{Fig2}) for short times. 
\\
We see that the force shows characteristic Drexhage-type oscillations. Remarkably we note that the force
 can change sign depending on the distance between the emitter and the surface.
This shows the importance of radiation modes to describe this effect.
 In fact, guided modes are excited strongly only in one direction giving a lateral force of constant sign. 
 
Another interesting feature of this force is that it is not conservative. In fact if we take the curl of the total force we obtain a non-vanishing result:
 \begin{multline}
 \nabla  \times \mathbf{F} = \frac{\partial {\operatorname{F}}_x}{\partial {\operatorname{z}}_{\operatorname{A}}}\hat y =\frac{{\operatorname{d}}^2}{\varepsilon _0}\bigg\{ \left( \frac{4\pi ^2}{\lambda _{10}^3{\operatorname{z}}_{\operatorname{A}}^2} - \frac{3}{\lambda _{10}{\operatorname{z}}_{\operatorname{A}}^4} \right)\cos\left( \frac{4\pi {\operatorname{z}}_{\operatorname{A}}}{\lambda _{10}} \right)\\
  + \left( \frac{3}{4\pi {\operatorname{z}}_{\operatorname{A}}^5} - \frac{5\pi }{\lambda _{10}^2{\operatorname{z}}_{\operatorname{A}}^3} \right)\sin\left( \frac{4\pi {\operatorname{z}}_{\operatorname{A}}}{\lambda _{10}} \right) \bigg\}\hat y \,.
 \end{multline}
 Hence the Casimir-Polder force is not conservative and can not be generally  derived from a potential, by just taking its gradient.  Such class of forces, called
 curl forces, have been explored in the past \cite{berry}.

We next consider the case the case of a dissipative dielectric medium described by a complex relative permittivity. An analytical expression can be derived in the non-retarded regimes when the distance between the emitter and the surface is much smaller than the atomic wavelength ${\operatorname{z}}_{\operatorname{A}} \ll  \lambda_{10}$ :
\begin{equation}
{\operatorname{F}}_x^{\text{nret}}(\textbf{r}_{\operatorname{A}},t) =
-\text{e}^{-\Gamma t}  \frac {3 {\operatorname{d}}^2}{8 \pi \varepsilon_0}
\frac {1}{{\operatorname{z}}^4_{\operatorname{A}}}
\frac {\text{Im}\, \varepsilon}{ \big| \varepsilon+1 \big|^2}  \; ,
\end{equation}
where $\epsilon=\epsilon(\omega_{10})$ and we have used the following Fresnel coefficient for p-polarised waves $r_p = \left( \varepsilon  - 1 \right)/\left( \varepsilon  + 1 \right)$. 
It shows a divergence when the emitter approaches the surface because of the lossy nature of the medium. Note that an analogous divergence
is observed in the spontaneous emission of a real dipole moment near a lossy surface, where the emitter is excited with a linear-polarised resonant laser beam \cite{sheel2,yeung}.

In the retarded regime, namely when the distance between the emitter and the surface is much greater than the atomic wavelength $z_A \gg \lambda_{10}$ we have:
\begin{multline}
{\operatorname{F}}_x^{\text{ret}}(\textbf{r}_{\operatorname{A}},t) =\text{e}^{-\Gamma t}\frac {{\operatorname{d}}^2 \pi}{ \varepsilon_0 \lambda_{10}^2} 
\frac {1}{{\operatorname{z}}_{\operatorname{A}}^2} \Big\{
\text{Re}\left[\frac {\sqrt{\varepsilon}-1}{\sqrt{\varepsilon}+1}
\right]\\
\times  \sin\left(4\pi \frac {{\operatorname{z}}_{\operatorname{A}}}{\lambda_{10}} \right) 
+ \text{Im}\left[\frac {\sqrt{\varepsilon}-1}{\sqrt{\varepsilon}+1}
\right] \cos\left(4\pi \frac {{\operatorname{z}}_{\operatorname{A}}}{\lambda_{10}}
\right)\Big\} \; ,
\end{multline}
where the following Fresnel coefficient for p-polarised wave has been used: $r_p = \left( \sqrt \varepsilon   - 1 \right)/\left( \sqrt \varepsilon   + 1 \right)$.

   The distance between two roots of the retarded case is a quarter of the emission
wavelength. This agrees with intuition. The atom emits at some time an
electromagnetic field while having a specific dielectric moment orientation.
The field is partially reflected by the surface and interacts with the dipole after some delay, when the dipole
moment has a different orientation. If the distance between the emitter and the surface is incremented by $\lambda_{10}/4$ the delay is incremented by $\Delta t=\lambda_{10}/2c$, leading to an opposite orientation of the dipole $\textbf{d} \to \textbf{d} \text{e}^{\text{i}\omega_{10} \Delta t}=- \textbf{d}$. This translates to a change of sign of the force. 

\begin{figure}[h!]
\centering
\includegraphics[width=9cm]{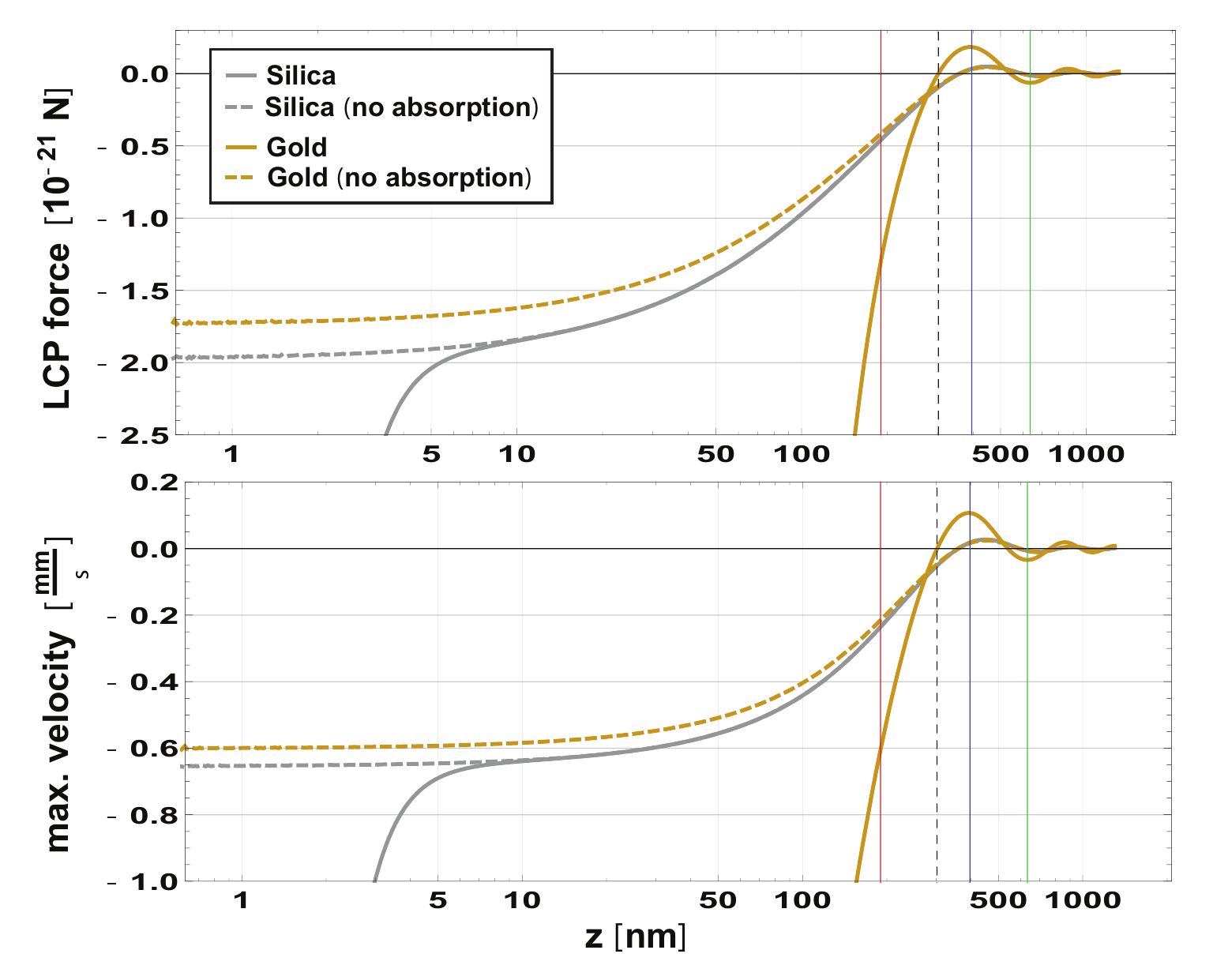}
\caption{Lateral CP force at initial time and long-time lateral velocity for an atom above silica and gold surfaces. Four vertical lines roughly depict characteristic points: the close distance regime (red, 190nm), vanishing (dashed black, 302nm), maximum (blue, 396nm) and minimum (green, 635nm) lateral CP force.}
\label{Fig3}
\end{figure}

In Fig. \ref{Fig3} we display the lateral force at the initial time when the emitter is near a gold dielectric ($\varepsilon_{Au} \approx 1.40 + 1.35 \text{i}$) or silica ($\varepsilon_{Si} \approx 1.45 + 2.05 \times 10^{-7}\text{i}$). 
The force shows a strong increment in the near-field regime, having a
magnitude of around $10^{-21}$ Newton. The same orders of magnitude have been obtained
for a circular emitter near a nanofiber \cite{LCPExp}.
To judge the strength of the effect we consider the lateral velocity acquired by the atom due to the recoil, which is in the $\mathrm{mm}/\mathrm{s}$ range.

The atom in the ground-state can be excited also to the hyperfine level
$\left| 6^2 P_{3/2},F' = 5,M'_F = -5 \right\rangle $ by using a left-handed circularly polarised laser beam.
The atom will emit a $\sigma^-$ photon and since the directionality depends on the polarisation of the emitted light, the lateral force will change sign.
This effect provides the opportunity to control the direction of the force changing the polarization of
the illuminating light or in other words the quantum state of the emitter.

Note that for all materials considered so far, the force is directed in the $- \hat x$ direction in the near-field regime. This is a consequence
of the right-hand rule in the lateral forces, already considered in the literature \cite{Kalhor,Mechelen}. According to this rule the direction of the lateral force, the decay direction of the evanescent wave ($\hat z$ direction), and the atomic spin ($\hat y$ direction) follows a right-hand rule. This means that the lateral force is directed along the $-\hat x $ direction, see Fig. \ref{fig1}. 

\section{Emission spectrum}
In this section we develop more closely the idea that the lateral force is a photon-recoil force.
For this reason we consider the emission spectrum:
\begin{equation}
\bar \Gamma ({\operatorname{z}}_{\operatorname{A}},\varphi ) = \int\limits_0^\infty  \text{d} k^\parallel  k^\parallel\hbar k^\parallel \gamma \left( {\operatorname{z}}_{\operatorname{A}},\mathbf{k}^\parallel  \right) \,,
\end{equation}
which describes the scattering of a momentum $\hbar k^\parallel$ along a particular direction parallel to the surface.
Taking into account formula Eq. (\ref{gamma4}) we find the emission spectrum is in general asymmetric:
\begin{multline}
\bar \Gamma ({\operatorname{z}}_{\operatorname{A}},\varphi ) = A\left( {\operatorname{z}}_{\operatorname{A}} \right) + B\left( {\operatorname{z}}_{\operatorname{A}} \right)\cos \varphi  + C\left( {\operatorname{z}}_{\text{A}} \right)\cos ^2\varphi\\  + D\left( {\text{z}}_{\text{A}} \right)\sin ^2\varphi \,,
\end{multline}
where:
\begin{align}\nonumber
A\left( {\text{z}}_{\text{A}} \right)= & - \frac{{\text{d}}^2}{4\pi^2 \varepsilon _0} \text{Re} \left[ \int\limits_0^\infty  \text{d} k^\parallel \frac{k^{\parallel 4}}{k^\bot} \text{e}^{2\text{i} k^ \bot  {\text{z}}_{\text{A}}} r_p \right],\\\nonumber
B\left( {\text{z}}_{\text{A}} \right)= &  \frac{{\text{d}}^2}{2\pi^2 \varepsilon _0} \text{Im} \left[ \int\limits_0^\infty  \text{d} k^\parallel k^{\parallel 3} \text{e}^{2\text{i} k^ \bot  {\text{z}}_{\text{A}}} r_p \right],\\\nonumber
C\left( {\text{z}}_{\text{A}} \right)= &  \frac{{\text{d}}^2}{4\pi^2 \varepsilon _0} \text{Re} \left[ \int\limits_0^\infty  \text{d} k^\parallel k^{\parallel 2} k^\bot \text{e}^{2\text{i} k^ \bot  {\text{z}}_{\text{A}}}r_p \right]\,,\\
D\left( {\text{z}}_{\text{A}} \right)= & - \frac{{\text{d}}^2}{4\pi^2 \varepsilon _0} \text{Re} \left[ \int\limits_0^\infty  \text{d} k^\parallel \frac{k^{\parallel 2}}{k^\bot} \frac{\omega^2}{c^2} \text{e}^{2\text{i} k^ \bot  {\text{z}}_{\text{A}}} r_s \right].
\end{align}
The emission is in general asymmetric along the $x$ direction  because of the factor $B\left( {\text{z}}_{\text{A}} \right)$:
\begin{multline}
\int\limits_{ - \pi /2}^{\pi /2} \text{d} \varphi \bar \Gamma ({\text{z}}_{\text{A}},\varphi ) - \int\limits_{\pi /2}^{3\pi /2} \text{d} \varphi \bar \Gamma ({\text{z}}_{\text{A}},\varphi ) = 4B\left( {\text{z}}_{\text{A}} \right)\\
= \frac{2{\text{d}}^2}{\pi^2 \varepsilon _0} \text{Im} \left[ \int\limits_0^\infty  \text{d} k^\parallel k^{\parallel 3} \text{e}^{2\text{i} k^ \bot  {\text{z}}_{\text{A}}} r_p \right]\,,
\end{multline}
The comparison with Eq. (\ref{eqn3}) shows that if the emission is asymmetric along the $x$-direction then the lateral force is finite.
\begin{figure}[!htb]
   \begin{minipage}{0.48\textwidth}
     \centering
     \includegraphics[width=1.0\linewidth]{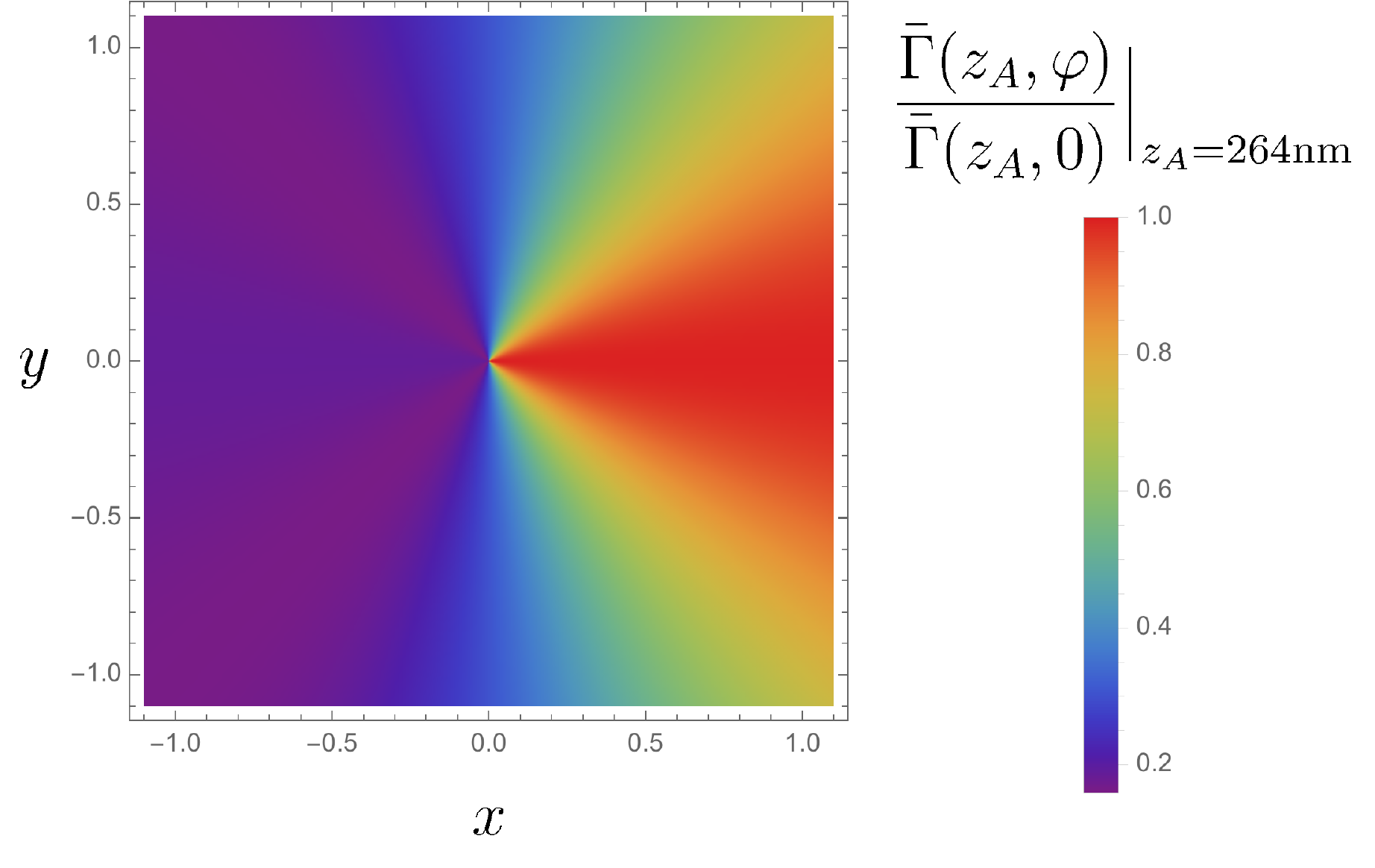}
     \caption{Asymmetric emission spectrum for ${\text{z}}_{\text{A}}=264 \, \rm{nm}$ along the $x$ direction.}\label{Fig:Data1}
   \end{minipage}\hfill
   \begin {minipage}{0.48\textwidth}
     \centering
     \includegraphics[width=1.0\linewidth]{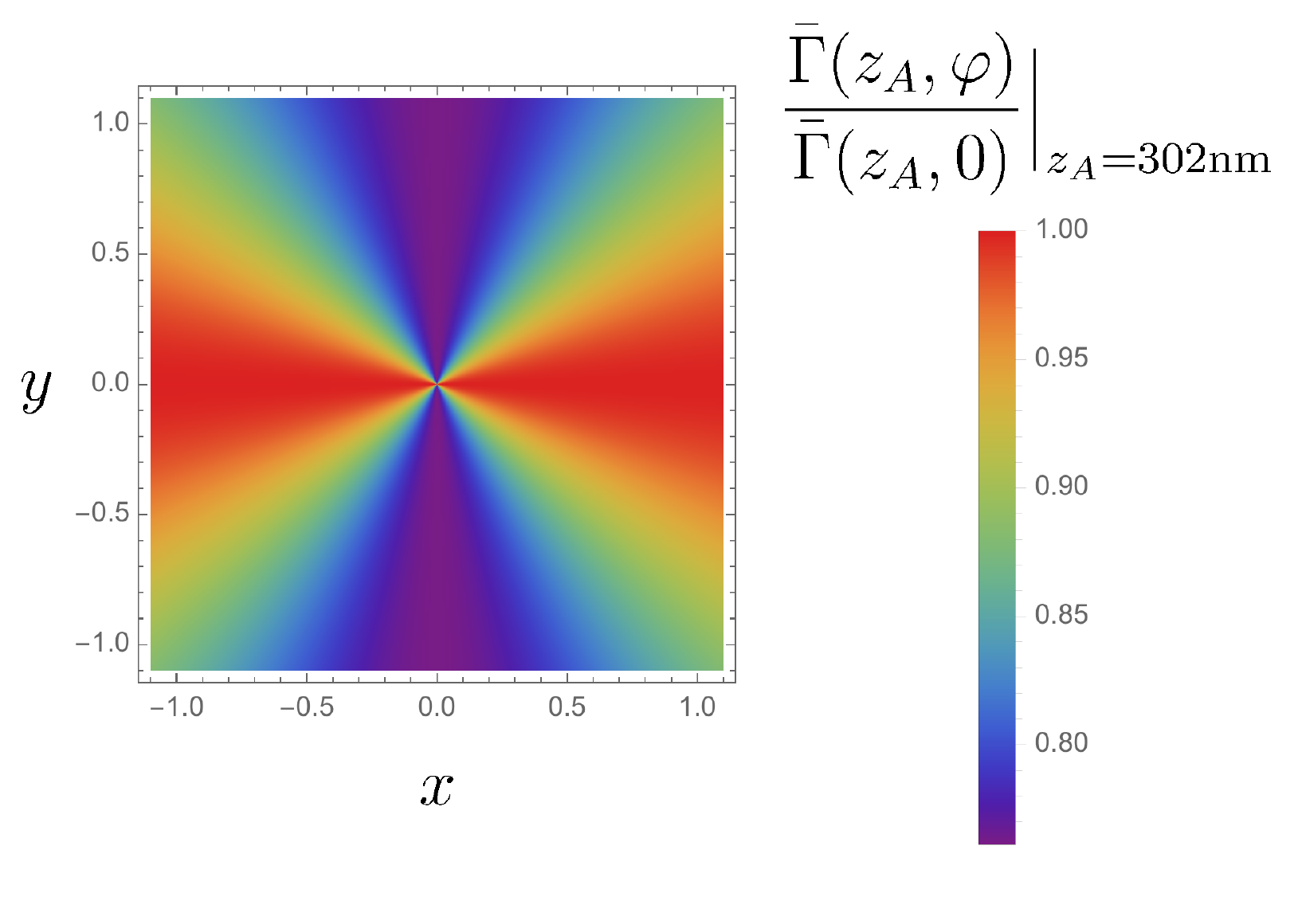}
     \caption{Symmetric emission spectrum for ${\text{z}}_{\text{A}}=302\, \rm{nm}$.}\label{Fig:Data2}
   \end{minipage}
\end{figure}\\
In Fig. \ref{Fig:Data1} and Fig. \ref{Fig:Data2} we show the emission spectrum for two atom-plate distances: ${\text{z}}_{\text{A}}=264\, \rm{nm}, 302\, \rm{nm}$. Figure \ref{Fig:Data1} shows an asymmetric emission with stronger emission in the positive $x$ direction; this suggests a negative lateral force as shown in Fig. \ref{Fig3}. For ${\text{z}}_{\text{A}}=302, \rm{nm}$ the emission spectrum is symmetric along the $x$-direction suggesting that the lateral force is zero, as shown in Fig. \ref{Fig3}.

\section{\label{sec:level1}Conclusion}
\noindent
We have predicted a  lateral Casimir-Polder force  for a circular excited emitter near a planar surface.
The underlying reason for a non-vanishing force is the breaking of time-reversal symmetry.
The sign of the force depends on the polarization of the emitted light and can be controlled by changing the quantum excited state of the emitter. We have also shown that the lateral force is an atom recoil force stemming from the asymmetric emission of the photon.  

Our dynamical approach of the field--atom coupling shows that the lateral force has a population-induced dynamics, decaying exponentially with time, on time scales of the inverse of the spontaneous decay rate. Moreover it exhibits characteristic Drexhage-type oscillations when changing the distance between the emitter and the surface. The near field regime is strongly influenced by the dissipative character of the medium: for short distances it converges to zero for a lossless medium, while it diverges if the medium has a  complex dielectric permittivity. 
This effect could be detecting measuring the velocity acquired by the atom after the recoil.

Our formalism which uses the scattering Green's tensor can be adopted for more complex geometries, like spheres, cylinders or resonating planar cavities which can enhance the spatial oscillations of the force.

\begin{acknowledgements}
We would like to thank M. Berry, F. J. Rodriguez-Fortu\~{n}o and S. Scheel for stimulating discussions.
ROW, PB and SYB are
grateful for support by the DFG (grants BU 1803/3-1 and GRK
2079/1) and the Freiburg Institute for Advanced Studies (S. Y .B).
\end{acknowledgements}

\end{document}